\documentclass[12pt]{article}
\usepackage{color}
\usepackage{latexsym}
\usepackage{epsfig,amssymb,euscript, mathrsfs}
\usepackage{amsmath}
\newsavebox{\ns}
\newsavebox{\dbrane}
\newsavebox{\dbshort}
\renewcommand{\arraystretch}{1.2}
\def\be{\begin{equation}}
\def\ee{\end{equation}}
\def\bea{\begin{eqnarray}}
\def\eea{\end{eqnarray}}

\newcommand{\nn}{\nonumber}

\newcommand\R{\mathbb{R}}

\newcommand\C{\mathbb{C}}
\newcommand\T{\mathbb{T}}
\newcommand\diff{\mathrm{d}}

\newcommand{\dd}{\mathrm{d}}

\newcommand{\ii}{\mathrm{i}}

\newcommand{\ex}{\mathrm{e}}

\newcommand{\rcut}{\varrho}

\newcommand\p{\mathcal{P}(p)}

\newlength{\sswidth}

\newcommand{\rc}{p_3}
\newcommand{\rd}{p_4}

\newcommand{\Anew}{A^\mathrm{nm}}

\numberwithin{equation}{section}       % equation numbers in each section

\begin{document}

\begin{titlepage}

\begin{center}

\today

\vskip 2.3 cm 

\vskip 5mm

{\Large \bf The gravity dual of supersymmetric gauge theories\\[3mm]

on a two-parameter deformed three-sphere}

\vskip 15mm

{Dario Martelli and  Achilleas Passias}

\vskip 1cm

\textit{Department of Mathematics, King's College London, \\
The Strand, London WC2R 2LS,  United Kingdom\\}

\end{center}

\vskip 2 cm

\begin{abstract}
\noindent  
We present  rigid  supersymmetric backgrounds for three-dimensional ${\cal N} = 2$ 
supersymmetric gauge theories, comprising a two-parameter $U(1)\times U(1)$-invariant 
deformed three-sphere, and  their gravity duals. 
These are described by supersymmetric solutions of four-dimensional ${\cal N}=2$ 
gauged supergravity with a self-dual metric on the ball and different instantons for the graviphoton field. 
We find two types of solutions, distinguished by their holographic free energies.
In one type the holographic free energy is constant, whereas 
in another type it depends in a simple way on the parameters and is generically complex. 
This leads to a conjecture for the localized partition function of a 
class of ${\cal N}=2$ supersymmetric gauge theories on these backgrounds.
\end{abstract}

\end{titlepage}

\pagestyle{plain}
\setcounter{page}{1}
\newcounter{bean}
\baselineskip18pt
\tableofcontents

\section{Introduction and summary}
\label{introsumma}

Localization techniques allow one to perform exact non-perturbative computations in 
supersymmetric field theories defined on a curved Euclidean manifold, 
thus motivating  the systematic study of rigid supersymmetry in curved space. In three dimensions the conditions for unbroken supersymmetry for ${\cal N}=2$ supersymmetric
field  theories  in Euclidean signature have been studied in  \cite{Klare:2012gn,Closset:2012ru}
following the approach of \cite{Festuccia:2011ws}. In these references it was shown that 
a supersymmetric Lagrangian can be constructed if there exists 
a spinor $\chi$ or $\tilde \chi$ obeying one of the following equations
\begin{equation}\label{mainrigid}
  \begin{split}
\nabla^{(3)}_\alpha \chi- \ii (A^{(3)}_\alpha  + V^{(3)}_\alpha)\chi + \tfrac{1}{2}H \gamma_\alpha \chi + \epsilon_{\alpha\beta\rho}V^{(3)}{}^\beta\gamma^\rho\chi  & =  0 ~,\\[2mm]
\nabla^{(3)}_\alpha \tilde \chi + \ii (A^{(3)}_\alpha  + V^{(3)}_\alpha)\tilde \chi + \tfrac{1}{2}H \gamma_\alpha \tilde \chi -\epsilon_{\alpha\beta\rho}V^{(3)}{}^\beta\gamma^\rho\tilde \chi & =  0 ~,
\end{split}
\end{equation}
arising from the rigid limit \cite{Festuccia:2011ws} of three-dimensional
new minimal \cite{Sohnius:1981tp}  supergravity\footnote{The fields $A^{(3)}$ and $V^{(3)}$ 
are related to those appearing in \cite{Klare:2012gn,Closset:2012ru},  
as $A^{(3)} = \Anew - \tfrac{3}{2} V^{\mathrm{nm}}$ and $V^{(3)}= \tfrac{1}{2} V^\mathrm{nm}$. The combinations
we use arise naturally from the point of view of four-dimensional gauged supergravity.}.
  The background fields consist of a metric $g_{\alpha\beta}$, 
an Abelian gauge field  $ A^{(3)}_\alpha$ coupling to the $U(1)_R$ current, a second vector 
field $V^{(3)}_\alpha$ obeying $\nabla^\alpha V^{(3)}_\alpha=0$,
and a scalar field $H$ \cite{Klare:2012gn,Closset:2012ru}. In Euclidean signature all fields are in principle 
complex, except for the metric which is usually required to be real.
Furthermore,  in general the spinor $\tilde\chi$ is not the charge conjugate of $\chi$ \cite{Festuccia:2011ws}.

Examples of geometries obeying equations (\ref{mainrigid}) were constructed in  \cite{Hama:2011ea} and \cite{Imamura:2011wg}, 
before the systematic analysis of \cite{Klare:2012gn,Closset:2012ru}. In particular, a $U(1)\times U(1)$-symmetric background, comprising  a one-parameter 
squashed three-sphere, was presented in  \cite{Hama:2011ea}.  The metric may be written as
\bea
\label{HHL1metric}
\diff s^2_3 & = & f^2(\vartheta)\diff \vartheta^2 + \cos^2\vartheta \diff \varphi_1^2 + \frac{1}{b^4}\sin^2\vartheta \diff \varphi_2^2~,
\eea
where $\vartheta \in [0,\tfrac{\pi}{2}]$, $\varphi_1,\varphi_2 \in [0,2\pi]$,  $b > 0$ is a constant squashing parameter,
and\footnote{The function originally used in  \cite{Hama:2011ea} is $f^{2}(\vartheta) = \sin^2\vartheta + b^{-4}\cos^2\vartheta$. 
However, it was later shown in \cite{Martelli:2011fu} that $f(\vartheta)$ can be an arbitrary function, provided it gives rise to a smooth metric
with the topology of the three-sphere. The specific choice presented in the text arises from the supergravity solution in \cite{Martelli:2011fu}.} 
$f^{-2}(\vartheta) = \sin^2\vartheta + b^4\cos^2\vartheta$. The  other background fields  are summarised in Table \ref{elliptictable}.
\begin{table}[h!]
\begin{center}
\renewcommand{\arraystretch}{2}
 \begin{tabular}{|c||c|c|c|}
 \hline
\small Background fields  & $H$ & $A^{(3)}$  & $V^{(3)}$ \\ \hline \hline
Ellipsoid \cite{Hama:2011ea} & 
$\displaystyle  -\frac{\ii}{f(\vartheta)}  $       & 
$\displaystyle  \frac{1}{2f(\vartheta)}\left(\diff \varphi_1 - \frac{1}{b^2}\diff \varphi_2 \right) $ & 
~$\displaystyle 0$~  \\ [.2cm] 
 \hline
 \end{tabular}
\end{center}
\caption{The $U(1)\times U(1)$-symmetric background.}
\label{elliptictable}
\end{table}

Two different  $SU(2)\times U(1)$-symmetric backgrounds, comprising  a biaxially squashed three-sphere, were 
presented in  \cite{Hama:2011ea} and \cite{Imamura:2011wg}, respectively. In both cases the metric may be written as
\bea
\diff s^2_3 & = & \diff \theta^2 + \sin^2\theta \diff \phi^2  + \frac{1}{v^2}(\diff  \psi + \cos\theta \diff \phi)^2~,
\label{fametric}
\eea 
where $\theta, \phi, \psi$ are standard Euler angles on $S^3$, so that  $\theta\in [0,\pi]$,  $\phi\in [0,2\pi]$, 
$ \psi \in [0,4\pi]$ and $v > 0$ is a constant squashing parameter. The rest of the background fields 
in the two cases\footnote{In the $\tfrac{1}{4}$ BPS case, using $\gamma_3\chi=-\chi$,   equation (\ref{mainrigid})  
is also solved by $V^{(3)}=0$, $H=+\tfrac{\ii}{2v}$, with the same metric and $A^{(3)}$ \cite{Hama:2011ea}. This is equivalent to using the shift
symmetry in eq. (4.2) of \cite{Closset:2012ru}, with $\kappa = \tfrac{\ii}{v}$, and noting that $A^{(3)} = \Anew - \tfrac{3}{2} V^{\mathrm{nm}}$ is invariant under this shift.}
are summarised in Table \ref{biaxialtable}, where 
 $\sigma_3 = \diff \psi + \cos\theta \diff \phi$.
\begin{table}[h!]
\begin{center}
\renewcommand{\arraystretch}{2}
 \begin{tabular}{|c||c|c|c|}
 \hline
\small Background fields  & $H$ & $A^{(3)}$  & $V^{(3)}$ \\ \hline \hline
$\displaystyle \frac{1}{4}$ BPS  \cite{Hama:2011ea} & 
$\displaystyle -\frac{\ii}{2v}$       & 
$\displaystyle \frac{1}{2v^2}(v^2-1)\sigma_3$ & 
$\displaystyle \frac{1}{2v^2}\sigma_3$ \\  [.2cm] \hline
$\displaystyle \frac{1}{2}$ BPS  \cite{Imamura:2011wg} & 
$\displaystyle -\frac{\ii}{2v}$     & 
$\displaystyle -\frac{1}{2v^2}\sqrt{1-v^2} \, \sigma_3$ & 
$\displaystyle \frac{1}{2v^2}\sqrt{1-v^2} \, \sigma_3$  \\ [.2cm] 
 \hline
 \end{tabular}
\end{center}
\caption{The two $SU(2)\times U(1)$-symmetric backgrounds.}
\label{biaxialtable}
\end{table}

In all  cases the partition function of an ${\cal N}=2$ supersymmetric gauge 
theory defined on these backgrounds can be computed exactly using localization, and reduces to a matrix model involving the 
double sine function $s_\beta(z)$, where $\beta $ is identified  with the parameter $b$ or 
related\footnote{In the $\tfrac{1}{2}$ BPS case the relation is $\tfrac{2}{v}= \beta + \tfrac{1}{\beta}$. In the $\tfrac{1}{4}$ BPS case, 
the partition function is identical to the round three sphere case, with all other background fields set to zero.} to the parameter $v$, respectively.

If a supersymmetric field theory defined on (conformally) flat space admits an AdS dual,
it is natural to ask whether a gravity dual still exists, when the field theory is 
placed on a non-trivial curved background. Gravity  solutions dual to gauge theories defined on the 
backgrounds discussed above were constructed in \cite{Martelli:2011fu}, \cite{Martelli:2011fw}, and \cite{Martelli:2012sz}, 
respectively. These are  described by one-parameter supersymmetric solutions of 
four-dimensional ${\cal N}=2$  gauged supergravity comprising a self-dual Einstein metric and an instanton 
for the graviphoton field. In particular, the metrics in these references were 
Euclidean versions of AdS$_4$ and Taub-NUT-AdS$_4$, which can be thought of as 
metrics on the ball. The holographic free energies
of these solutions were shown to agree with the leading large $N$ free energy of the field theory, defined as 
minus the logarithm of the localized partition function.

In this paper  we discuss a family of rigid supersymmetric backgrounds 
on the three-sphere depending on \emph{two parameters}, together with its gravity dual. 
Namely, an  asymptotically locally Euclidean AdS metric, whose  conformal structure at infinity reproduces the fields
$g_{\alpha\beta},A^{(3)}_\alpha,V^{(3)}_\alpha,H,\chi, \tilde \chi$, obeying equation (\ref{mainrigid}). Here we present 
the three-dimensional boundary data, leaving the discussion of the bulk solution and its properties to the central part of the paper. 
The  metric may be written (up to an irrelevant overall factor) as 
\bea
\diff s^2_3 &=& \frac{\diff\theta^2}{f(\theta)} + f(\theta) \sin^2\theta \, \diff \hat \phi^2
+(\diff\hat\psi + (\cos\theta+ a\sin^2\theta)\diff\hat \phi)^2~,
\label{finalmetric}
\eea
where $\theta \in [0,\pi]$  and
\bea 
f(\theta) & =  & v^2 - a^2 \sin^2\theta - 2  a \cos\theta~.
\eea
The angular variables $\hat \phi$, $\hat \psi$ 
do not have canonical periodicities, and in particular the two-dimensional ``transverse'' 
metric is not globally well-defined.  The global structure of the metric is elucidated introducing two angular coordinates as
\bea
\hat \psi \, = \, \frac{1}{v^2-2a}\varphi_1   + \frac{1}{v^2 +2 a}\varphi_2 ~,\qquad \hat \phi \, =\,   -  \frac{1}{v^2-2a}\varphi_1  +  \frac{1}{v^2 +2 a}\varphi_2~,
\eea
where $\varphi_1,\varphi_2\in [0,2\pi]$. The metric in the coordinates $\theta,\varphi_1,\varphi_2$ describes a smooth three-sphere, viewed as a 
$\T^2_{\varphi_1,\varphi_2}$ fibered over the interval $[0,\pi]$.  The two (real) parameters are  $v>0$ and $a$, with $2|a|<v^2$.
The remaining background fields are given by
\bea
H  &=&   \ii (\tfrac{1}{2}- a\cos\theta) ~,\nn\\
A^{(3)} &  = & Q  (\diff \hat \psi +  \cos\theta \diff \hat \phi ) ~, \label{finalbacks}\\
 V^{(3)} & = & \frac{v^2-1}{4Q} (\diff\hat \psi+(\cos\theta+ a\sin^2\theta)\diff\hat \phi) ~,\nn
 \eea
where $Q=Q(a,v)$ depends on the two parameters. More precisely, for a fixed conformal class of metric, 
there is a discrete choice of $Q$, yielding different spinors\footnote{As we shall explain, for any value of $Q$, $-Q$ also yields a solution. We will denote
as $\chi (Q)$ the spinors for one choice of sign, and as $\tilde \chi (Q) =\ \chi (-Q)$, which formally solve the second equation in \eqref{mainrigid}. }; see (\ref{iandq}) below. Note that the fields 
$A^{(3)}$,  $V^{(3)}$ in \eqref{finalbacks} are \emph{globally} defined on the three-sphere and  $\nabla^{(3)}{}^\alpha V^{(3)}_\alpha=0$.

This family of backgrounds admits generically one Killing spinor $\chi$ and
includes all the previously known ones as special one-parameter families. 
One case is obtained by setting $v=1$, with $a\in [-\tfrac{1}{2},\tfrac{1}{2}]$. 
The resulting metric has still $U(1)\times U(1)$ isometry, and in fact is diffeomorphic to the  metric in (\ref{HHL1metric}) \cite{Martelli:2011fu}. 
Another case is obtained by setting $a=0$. The resulting metric is the biaxially squashed metric 
(\ref{fametric}) and there are two inequivalent choices of background fields, 
corresponding to the $\tfrac{1}{2}$ BPS and  
$\tfrac{1}{4}$ BPS backgrounds in Table \ref{biaxialtable}, respectively.

We shall see that these backgrounds arise at the boundary of a family of supergravity solutions comprising a self-dual Einstein metric on the ball and
different choices of instantons for the graviphoton field. We can then use these solutions to compute the holographic free energy in the various cases. 
This depends on the choice of background fields  and takes the remarkably simple form 
\bea
I \ = \ \frac{\pi}{2G_4}
\begin{cases}
1\\[2mm]
 \frac{1}{4} \left( \beta+ \frac{1}{\,\beta} \right)^2
 \end{cases}  \mathrm{for}\quad ~
Q\ = \ \mp \tfrac{1}{2} \begin{cases}  v^2-1  \qquad\qquad  \qquad~~~\,\mathrm{Type~I}\\[2mm] \sqrt{a^2 +1 - v^2} \pm a \qquad~ \mathrm{Type~II}
\end{cases},
\label{iandq}
\eea
where the parameter $\beta$ is defined through $Q   =  \frac{1}{2}\frac{\beta^2-1}{\beta^2+1}$.
 In the  solution of Type I the free energy takes the constant value of the round three-sphere, 
  independently of the parameters $a$ and $v$. This  may be regarded as a deformation 
of the $\tfrac{1}{4}$ BPS $SU(2)\times U(1)$-invariant background of  \cite{Hama:2011ea}.   
In the solutions of Type II the free energy  depends on the two parameters $a$ and $v$ only through $Q$, which generically takes 
\emph{complex} values.

In the remainder of the paper we derive the results previewed above. We start by constructing a family of Euclidean 
asymptotically locally AdS solutions of minimal gauged supergravity, reproducing the above 
backgrounds at their  conformal boundary. In section \ref{sugrasol} we first present the local form of the  solutions, and then
discuss their global properties.  Supersymmetry of the solutions will be demonstrated in section \ref{susysec}, where the explicit form of 
the Killing spinors in the bulk and on the boundary will be provided. In section \ref{parameters} we discuss the parameter space of 
solutions. In section \ref{holofree} we write the holographic free energy associated to the solutions. Section \ref{discussect} concludes with 
a brief discussion. In the appendices we derive two integrability results and give more details on the Killing spinors.

\section{Supergravity solutions}
\label{sugrasol}

The action for the bosonic sector of $d=4$, $\mathcal{N}=2$ gauged supergravity \cite{Freedman:1976aw} is
\bea\label{4dSUGRA}
S &=& -\frac{1}{16\pi G_4}\int \diff^4x\sqrt{\det g_{\mu\nu}}\left(R + 6g^2 - F_{\mu\nu}F^{\mu\nu}\right)~,
\eea 
where $R$ denotes the Ricci scalar of the four-dimensional metric $g_{\mu\nu}$, and the cosmological constant 
is given by $\Lambda=-3 g^2$. The graviphoton is an Abelian gauge field $A$ with field strength 
$F=\diff A$.  The equations of motion derived from (\ref{4dSUGRA}) are
\bea\label{EOM}
R_{\mu\nu} + 3g^2 g_{\mu\nu}  &=& 2\left(F_\mu^{\ \rho}F_{\nu\rho}-\tfrac{1}{4}F^2 g_{\mu\nu}\right)~,\nonumber\\
\diff *_4F &=&0~. 
\eea
Notice that when $F$ is self-dual the right hand side of the Einstein equation vanishes and 
(in Euclidean signature) the equations of motion are consistent with  a complex gauge field $A$, 
while the metric  $g_{\mu\nu}$ remains real. It was shown in \cite{Gauntlett:2007ma, Gauntlett:2009zw} that any solution to $d=4$, $\mathcal{N}=2$ 
gauged supergravity uplifts (locally) to a  solution of eleven-dimensional  supergravity.

Our starting point is the local form of a class of solutions  to (\ref{EOM}), originally found by
Plebanski-Demianski  \cite{Plebanski:1976gy}. These  are  the most general solutions of Petrov type D, and it is this property
that allows one to solve the equations in closed form. Many known solutions arise as particular limits of these, including the solutions 
presented in \cite{Martelli:2011fu} and \cite{Martelli:2011fw}, as we shall discuss in the course of the paper.
We will adopt  the form of the solutions essentially as presented in \cite{amo}. In Euclidean signature, the metric can be written as
\bea\label{PD}
\dd s^2 = \frac{\mathcal{Q}(q)}{q^2-p^2} (\diff \tau + p^2 \diff\sigma)^2 + \frac{q^2-p^2}{\mathcal{Q}(q)} \diff q^2
+ \frac{p^2-q^2}{\mathcal{P}(p)}\diff p^2 + \frac{\mathcal{P}(p)}{p^2-q^2}(\diff \tau + q^2 \diff\sigma)^2~,
\eea
where $\mathcal{P}(p)$  and $\mathcal{Q}(q)$ are quartic polynomials given 
by\footnote{To obtain the metrics in the Euclideanized form presented here, one should
take the solutions as presented in \cite{amo} and map $p\mapsto \ii p $,   $\tau \mapsto -\ii \tau$, $\sigma \mapsto -\ii \sigma$,  $N \mapsto \ii N$, $Q\mapsto \ii Q$ 
(and reverse the signature of the metric).
This yields the solution written  in Appendix A of \cite{Martelli:2011fu}, up to some sign differences in the parameters. In particular, the two Euclidean 
solutions are related by $E \mapsto - E$, $N\mapsto - N$, $M\mapsto - M$, $P\mapsto -P$.}
\bea
\mathcal{P}(p) &=&  g^2 p^4 + E p^2 - 2 N p - P^2 + \alpha ~,\nn\\
\mathcal{Q}(q) &=& g^2 q^4 + E q^2 - 2 M q - Q^2 + \alpha~.
\eea
Here  $E,M$ and $N$ are arbitrary real constants, while $P,Q$ and  $\alpha$ may be complex. Setting $g=1$ without loss of generality, the gauge field reads
\bea
A & =  & \frac{pP-qQ}{q^2-p^2}\diff\tau + pq\frac{qP-pQ}{q^2-p^2}\diff\sigma~,
\label{generalgaugefield}
\eea
and therefore it may take \emph{complex} values. 

In this paper we will be interested in the subset of these solutions that correspond to \emph{supersymmetric global}
metrics on the ball. Different topologies are certainly possible, but we will not discuss these 
here\footnote{See \cite{Behrndt:2002xm} for a discussion in a different context.}.  Requiring a regular metric with  
ball topology leads  to the condition $N=M$, implying the  metrics are Einstein with self-dual Weyl tensor, 
and hence  $F$ is an instanton. Supersymmetry of the solutions will be addressed in section  \ref{susysec}.

The metric (\ref{PD}) is highly symmetric in the $p,q$ variables. As we need a non-compact direction
we will take $q$ as a coordinate that goes to infinity. In particular, we can take 
$q \in [q_+,\infty]$ or $q \in [-\infty,q_-]$, 
where $q_+(q_-)$ is the largest (smallest) root of $\mathcal{Q}(q) = 0$. Then we have that  $\mathcal{Q}(q) \geq 0$ and positivity of the metric
requires that $p^2-q^2\leq 0$, and $\mathcal{P}(p) \leq 0$.  Therefore $p$ lies in a closed interval $p\in [p_-,p_+]$ where $p_-$ and $p_+$ are two adjacent 
real roots  of $\mathcal{P}(p) = 0$, that we also require to be \emph{simple}.
 Writing 
\bea
\mathcal{P}(p)  & = & (p-p_1)(p-p_2)(p-p_3)(p-p_4)~,
\eea
with  $p_1+p_2+p_3+p_4 = 0$, depending on the reality properties of the four roots we can consider two cases. If there are
 \emph{four real} roots, then we can introduce the ordering 
\bea
 p_1 \leq p_2 < p_3 < p_4~,
\eea
with $p_1<0$ and $p_4>0$, and without loss of generality we will take $p\in[p_3,p_4]$.  
If there are  \emph{two real and two complex roots}, then we denote $p_3=p_-, p_4=p_+ \in \mathbb{R}$, with  $p_4>0$, and  $p_1,p_2=(p_1)^* \in \mathbb{C}$. 
In addition, Re$[p_1] = -\frac{1}{2}(p_3+p_4)$.

The regularity analysis is divided in two parts. We will first  address regularity of the metric at the boundary, where $|q|\to \infty$,  
and then regularity of the metric in the interior.  We will follow \cite{Martelli:2007pv}, where a similar analysis was performed.

\subsection{Regularity on the boundary}
\label{boundreg}

We start by demanding that the boundary metric has the topology of a three-sphere. 
 Specifically, we take $q \in [q_+,\infty]$ so that  for $q \to \infty$ the metric becomes
\bea
\dd s^2 \ \simeq \ \frac{\diff q^2}{q^2} + q^2 \diff s^2_3~,
\eea
where the boundary metric is 
\bea
\label{bdrymetric}
\dd s^2_3 & = & -\frac{\dd p^2}{\p}  - \p \dd\sigma^2 + (\dd\tau + p^2 \dd\sigma)^2~,
\eea
and recall that  $\mathcal{P}(p) \leq 0$ for $p \in [\rc,\rd]$. 
We can analyse regularity of the metric (\ref{bdrymetric}) by studying the vanishing loci of a generic Killing vector 
\bea
k \ = \ a \, \partial_\tau + b \, \partial_\sigma~,~~~~~a,b \in \mathbb{R}~,
\eea
where we can take $b \neq 0$ without loss of generality. 
The norm of $k$ is
\bea
\left\| k \right \|^2 \ = \ (a + b p^2)^2 - p^2\mathcal{P}(p)~,
\eea
which is a sum of two positive terms, and therefore it vanishes if and only 
if   ($p = \rc$, $a/b = - \rc^2)$ or ($p = \rd$, $a/b = -  \rd^2$).
Namely, we have the following two vanishing Killing vectors
\bea
V_1 \ = \ \rc^2 \partial_\tau - \partial_\sigma~,  ~~~~~   V_2 \ = \ \rd^2 \partial_\tau - \partial_\sigma~,
\eea
at $p=p_3$ and $p=p_4$ respectively. We can introduce coordinates along these two Killing vector fields defining
\bea
\tau &=& \rc^2 \phi_1 + \rd^2 \phi_2 ~,\nn \\
\sigma &=& - \phi_1 -\phi_2~,
\label{stchange}
\eea
so that $V_i  = \frac{\partial}{\partial\phi_i}$, $i  =  1,2$.
In terms of the new angular variables $\phi_1,\phi_2$ the boundary metric reads
\bea
\diff s^2_3 & = & - \frac{\diff p^2}{\mathcal{P}(p)} + [(p^2-\rc^2)^2 - \mathcal{P}(p)] \diff\phi_1^2 + [(p^2-\rd^2)^2 - \mathcal{P}(p)] \diff\phi_2^2 \nn \\
                   & + &  2[(p^2-\rc^2)(p^2-\rd^2) - \mathcal{P}(p)] \diff\phi_1\diff\phi_2~.
                   \label{fiufiu}
\eea
We now proceed by studying the behavior of the metric near to the end-points of the interval $[\rc,\rd]$. 
Near to $p = \rc$, setting  $p = \rc - \tfrac{\mathcal{P}'(p_3)}{4}r^2$, 
at first order in $r^2$  the metric takes the form 
\bea
\diff s^2_3 \ \simeq \ \diff r^2 + r^2 \frac{\mathcal{P}'(\rc)^2}{4}(\diff\phi_1 + c_- \diff\phi_2)^2 + (\rc^2-\rd^2)^2 \diff\phi_2^2~,
\label{near3}
\eea
where $c_-$ is constant whose value is irrelevant\footnote{We have $c_-   =  1 - \frac{2(\rc^2-\rd^2)\rc}{\mathcal{P}'(\rc)}$ and $c_+   =  1 - \frac{2(\rd^2-\rc^2)\rd}{\mathcal{P}'(\rd)}$.}. Similarly,  near to $p=\rd$ 
setting  $p = p_4 - \tfrac{\mathcal{P}'(p_4)}{4}r^2$, 
at first order in $r^2$  the metric takes the form 
\bea\label{near2}
\diff s_3^2 \ \simeq \ \diff r^2 + r^2 \frac{\mathcal{P}'(\rd)^2}{4}(\diff\phi_2 + c_+ \diff\phi_1)^2 + (\rc^2-\rd^2)^2 \diff\phi_1^2~.
\eea
Finally, defining 
\bea\label{var1}
\varphi_1 \ = \ \frac{\mathcal{P}'(\rc)}{2} \phi_1~,~~~~~\varphi_2 \ = \ \frac{\mathcal{P}'(\rd)}{2}\phi_2~,
\eea
where $\varphi_1,\varphi_2$ have period $2\pi$,  near to each root $p=\rc$ and $p=\rd$, the metrics (\ref{near3}) and 
(\ref{near2}) 
describe  smooth $\R^2$ fibrations over a circle. 
Equivalently, the space can be viewed as a 
 $\T^2$ fibration over the interval, where one of the cycles of the torus collapses smoothly at each end-point of the interval. 

In summary, we have shown that for \emph{any value} of the parameters of the solutions, there is a choice of periodicities of the angular coordinates, such that 
the boundary is topologically a three-sphere.

\subsection{Regularity  in the bulk}
\label{bulkreg}

Next we study regularity of the metric in the interior. Without loosing generality we will take $q \in [q_+,\infty]$,
where $q_+$  is the largest root of $\mathcal{Q}(q)=0$ and
\bea
q \geq q_+ \geq p_4 > 0~.
\eea
The analysis then splits into various sub-cases. Firstly, we will check regularity at fixed  $q$ such that $q > q_+$. 
We will then study separately the collapse of the metric at $q=q_+$.

\subsubsection*{Regularity at $q=q_0 > q_+$}

We fix a value of $q=q_0$ and consider the induced metric
\bea
\dd s^2_{q=q_0} &=&  \frac{p^2- q_0^2 }{\p }\dd p^2   -\frac{\mathcal{Q}(q_0)}{p^2- q_0^2} (\dd\tau + p^2 \dd\sigma)^2 + \frac{\p}{p^2-q_0^2} (\dd\tau + q_0^2\dd\sigma )^2   
 ~.
\eea
To study the collapse of this metric near to the zeroes of $\mathcal{P}(p)$ we  change coordinates again as in \eqref{stchange}, so that the metric reads
\bea
\dd s^2_{q=q_0} & = & -\frac{q_0^2 - p^2}{\p} \dd p^2   + \frac{1}{q_0^2 -p^2}\left[ \mathcal{Q}(q_0) (p_3^2 - p^2)^2 - \p (p_3^2- q_0^2)^2\right]\dd \phi_1^2 \nn\\
&+& \frac{1}{q_0^2 -p^2}\left[ \mathcal{Q}(q_0) (p_4^2 - p^2)^2 - \p (p_4^2- q_0^2)^2\right]\dd \phi_2^2\nn\\
& + & \frac{2}{q_0^2-p^2} \left[ \mathcal{Q}(q_0)  (p^2-p^2_3)(p^2-p_4^2)  - \p (q_0^2 - p_3^2)(q_0^2- p_4^2)  \right] \dd\phi_1 \dd \phi_2~. \nn
\eea
As before, near to $p=p_3$ setting $p=p_3- \frac{1}{4} \frac{\mathcal{P}'(p_3)}{q_0^2- p_3^2}r^2 $, 
 we can write the metric at leading order as 
\bea\label{nearp1}
\dd s_{q=q_0}^2 & \simeq & \dd r^2  + r^2  \frac{\mathcal{P}'(p_3)^2}{4}(\diff\phi_1 + \hat{c}_- \diff\phi_2)^2 + \hat{d}_- \dd \phi_2^2~,
\eea
where $\hat{c}_-, \hat{d}_-$ are constants depending on a fixed $q_0$, whose values are again irrelevant. 
This is automatically regular, given the periodicity of $\phi_1$ already fixed.  Of course, similarly, the metric is regular also at $p=p_4$.

\subsubsection*{Regularity at $q = q_+ = p_4$}

We begin by studying regularity of the metric near $q=q_+=p_4$ and $p=p_3$. Following \cite{Martelli:2007pv}, 
we introduce new   coordinates
\bea\label{coc}
R_1^2 &=& a_1(q - p_3)(p - p_3)~,\nn\\
R_2^2 &=& a_2(q-p_4)(p -p_4) ~,
\eea
where 
\bea
a_1 \ = \ -\frac{4(p_3+p_4)}{\mathcal{P}'(p_3)}~,\qquad 
a_2 \ = \ -\frac{4(p_3+p_4)}{\mathcal{Q}'(p_4)}~,.
\eea
Then  expanding near to $q=p_4$, $p=p_3$ (\emph{i.e.}  $R_1 = R_2=0 $)  the  metric becomes 
\bea
\diff s^2_{q\simeq p_4, \, p\simeq p_3}   &\simeq&  \diff R_1^2  +R_1^2 \diff \varphi_1^2 +\diff R_2^2 +  R_2^2 \frac{\mathcal{Q'}(p_4)^2}{\mathcal{P'}(p_4)^2}\diff \varphi_2^2~.
\eea
For this to be a smooth metric on $\R^4 = \R^2_1 \oplus \R^2_2$ we conclude that we must have $\mathcal{Q}'(p_4) =  \mathcal{P}'(p_4)$,
which in turn implies\footnote{We assume that $M=N \neq 0$. In the $M = N = 0$ case the bulk is Euclidean AdS$_4$.}
\bea
M  &= & N~.
\eea
Note that the Weyl tensor is self-dual precisely if and only if $N=M$.

Finally, we consider regularity of the metric near to $q = q_+ = p_4$ and $p=p_4$. Using $\mathcal{Q}'(p_4) = \mathcal{P}'(p_4)$ and changing coordinates as in 
\eqref{coc}, but with a different choice of constants $a_1,a_2$ given by
\bea
a_1 \ = \ \frac{8 p_4}{\mathcal{P}'(p_4)}~, \qquad a_2 \ = \ - \frac{8 p_4}{\mathcal{Q}'(p_4)}~,
\eea
we find the following expansion of the metric:
\bea
\diff s^2_{q\simeq p_4, \, p\simeq p_4} & \simeq & \diff R_1^2 + 
\left(\frac{1}{2} \frac{\mathcal{P'}(p_4)}{\mathcal{P}'(p_3)}\frac{p_3+p_4}{p_4} \right)^2 R_1^2 \diff\varphi_1^2 + \diff R_2^2 + R_2^2 \diff\varphi^2_2  ~,
\eea
which is a smooth metric on $\R^4 = \R^2_1 \oplus \R^2_2$, near to $R_1=$constant, $R_2=0$.

\subsection{Regularity of the gauge field}
\label{gaugereg}

We conclude this section discussing regularity of the gauge field.  The field strength of the gauge field (\ref{generalgaugefield}) 
is given by
\bea
F  & =  & \frac{Q(p^2+q^2) - 2 P p q}{(q^2-p^2)^2}\diff q \wedge (\diff\tau+p^2\diff\sigma) \nn\\[2mm]
& +& \frac{P (p^2+q^2) - 2 Q p q}{(q^2-p^2)^2} \diff p \wedge (\diff\tau+q^2\diff\sigma)~.
\eea
However, the condition $M=N$ implies that the metric is Einstein, 
and since the energy-momentum  tensor of $F$ is proportional to $P^2-Q^2$, the equations of motion (\ref{EOM}) 
imply that $P = \pm Q$, \emph{i.e.}  $F$ is either self-dual or anti-self-dual. The field strength then simplifies to
\bea
F  & =  & \frac{Q}{(q\pm p)^2} \left[\diff q \wedge (\diff\tau+p^2\diff\sigma) \pm \diff p \wedge (\diff\tau+q^2\diff\sigma)\right]~,
\eea
respectively. In order to have a non-singular $F$ we  need to 
make sure that  $p\pm q$ is never zero. However, the lower sign leads to a  zero at $p=q=q_4$, and therefore corresponds to a singular instanton. 
We then need to take $Q=P$, and to ensure that this  instanton is non-singular we must  impose the condition  $p_3+ p_4 >0$.
Changing  angular coordinates as in (\ref{stchange}) one checks  that the one-forms $\diff\tau+p^2\diff\sigma$ 
and $\diff\tau+q^2\diff\sigma$ are globally defined on $\R^4$, and hence the field strength is globally defined.

Finally, we note that upon using $Q=P$ the gauge field (\ref{generalgaugefield}) is  not well-defined 
at the end-points of the interval $p=p_3$ and $p=p_4$. This  can be easily corrected 
by adding a closed part to (\ref{generalgaugefield}), so that the total gauge field 
\bea
A_\mathrm{global} & =  & \frac{Q}{p+q} \left( \diff\tau + pq\diff\sigma \right) + \frac{Q}{p_3+p_4} (\diff \tau - p_3 p_4 \diff \sigma)~ ~,
\label{globala}
\eea
is now globally defined. In the next section, we will work with the singular gauge to begin with, explaining 
in the end how the global form \eqref{globala} affects the discussion of the Killing spinors.

\section{Supersymmetry}
\label{susysec}

We will now determine the subset of solutions that preserve supersymmetry, namely that admit at least one solution to 
the Killing spinor equation 
\bea\label{KSE}
\left[\nabla_\mu + \frac{1}{2}\Gamma_\mu - \ii A_\mu + \frac{\ii}{4}F_{\nu\rho}\Gamma^{\nu\rho}\Gamma_\mu\right]\epsilon &=& 0~
\eea
of four-dimensional minimal gauged supergravity. Here $\epsilon$ is a Dirac spinor and $\Gamma_\mu$, $\mu=1,\dots,4$, generate the Clifford
algebra Cliff$(4,0)$, so $\{ \Gamma_\mu,\Gamma_\nu\}=2g_{\mu\nu}$, where  $g_{\mu\nu}$ is our (real) Euclidean metric. However, we allow 
the gauge field $A_\mu$ to be complex.

In \cite{amo} the authors studied which of the Plebanski-Demianski solutions are supersymmetric solutions of $d=4, \mathcal{N}=2$ gauged 
supergravity and  derived a set of \emph{necessary} conditions  on the 
parameters\footnote{The sufficiency of these conditions was recently studied in \cite{Klemm:2013eca} in Lorentzian signature. 
However, Euclidean self-dual solutions do not have a Lorentzian origin, therefore the analysis of \cite{Klemm:2013eca} does not apply to the solutions of interest to us.},  
whose appropriate Euclidean version reads:
\bea
MP - NQ  &=& 0~, \nonumber\\
(N^2 -M^2 +E(P^2-Q^2))^2   & = &  4\alpha (P^2-Q^2)^2~. 
\label{BPSAMO}
\eea
For $M=N=0$ these yield $E^2 =4\alpha$ and no constraints on $P$ and $Q$.  However $M=N=0$ implies the metric is Euclidean 
AdS$_4$ and hence the gauge field must be an instanton \emph{i.e.}\ $P=Q$    \cite{Martelli:2011fu}. In our analysis we will  assume
$M=N \neq 0$, so that  again we must have $P=Q$, and in the limit $M\to 0$ our conclusions will agree with the $M=0$ case.  To summarise, so far 
the number of independent parameters has been reduced to four, for example $M,E,Q,\alpha$. However, these solutions do not  preserve supersymmetry, unless a further condition 
is satisfied. Below we determine this condition, and using this we present the explicit solution to \eqref{KSE}.

We have found convenient to start by deriving the asymptotic form of the  Killing spinor equation, which will be satisfied by a spinor $\chi$ defined on the three-dimensional boundary.  
We define the orthonormal frame
\bea
\label{4dframe}
&&e^1 = \sqrt{\frac{p^2-q^2}{\mathcal{P}(p)}} \diff p ~,\qquad \qquad \qquad \, e^2 = \sqrt{\frac{\mathcal{P}(p)}{p^2-q^2}}(\diff\tau+q^2\diff\sigma) ~,\nonumber\\[2mm]
&&e^3 = \sqrt{\frac{\mathcal{Q}(q)}{q^2-p^2}}(\diff\tau+p^2\diff\sigma)~, \qquad  e^4 = \sqrt{\frac{q^2-p^2}{\mathcal{Q}(q)}} \diff q ~,
\eea
and adopt the following representation of the gamma matrices
\bea
\hat{\Gamma}_a &=& \left(\begin{array}{cc}0 & \sigma_a \\ \sigma_a & 0\end{array}\right)~,\qquad \hat{\Gamma}_4 \ = \ \left(\begin{array}{cc}0 & \ii \mathbb{I}_2 \\ -\ii \mathbb{I}_2 & 0\end{array}\right)~,
\eea
where $\sigma_a$ are the Pauli matrices. Expanding the $q$ component of  (\ref{KSE}) for large $q$ we obtain the following 
 asymptotic form of the Killing spinor
\bea
\label{expaspinors}
\epsilon & = & \left( \begin{array}{c} \epsilon_+ \\ \epsilon_- \end{array}\right) \, = \,
 \left( \begin{array}{c}  -q^{1/2}  \left[1  -    \frac{1}{2q }\left(p  - \frac{M}{Q} \sigma_3\right) \right] \ii \chi \\[1.8mm]
~\, q^{1/2}  \left[1+   \frac{1}{2q}\left(p  - \frac{M}{Q} \sigma_3\right)  \right] \chi \end{array}\right) + {\cal O}(q^{-3/2})~.
\eea
In particular, we find that $\chi$ satisfies the equation
\bea
\label{kseboundary}
\left[ \nabla^{(3)}_\alpha - \ii (A^{(3)}_\alpha  + V^{(3)}_\alpha) + \ii \frac{p}{2} \, \gamma_\alpha + \epsilon_{\alpha\beta\rho} V^{(3)}{}^\beta \gamma^\rho\right] \, \chi & = & 0~,
\eea
where $\gamma_\alpha$, $\alpha = 1,2,3$ are the Pauli matrices,
$\nabla^{(3)}_\alpha$ denotes the covariant derivative with respect to the metric (\ref{bdrymetric}) and
 \bea
A^{(3)} \,= \, Q p \diff\sigma~,\quad ~~~~~~~V^{(3)} \, = \, \frac{M}{2Q} (\diff \tau + p^2 \diff \sigma) ~.
\label{aandv}
\eea
To write \eqref{kseboundary} we used the following three-dimensional orthonormal frame 
\bea\label{3dframe}
\hat e^1 \,=\, \frac{\diff p}{\sqrt{-\mathcal{P}(p)}}~,~~~~~\hat e^2 \,=\,  \sqrt{-\mathcal{P}(p)} \diff\sigma~,~~~~~
\hat e^3\,=\, \diff \tau + p^2 \diff \sigma~,
\eea
inherited from the four-dimensional frame (\ref{4dframe}). In Appendix \ref{integrab} 
we have studied  the  integrability condition for this equation and found that  this leads to the following equation for the parameters
\bea
4\alpha & = & \left(\frac{M^2}{Q^2}+E\right)^2 ~,
\label{cool}
\eea
that is independent of the BPS equations in   (\ref{BPSAMO}).  Interestingly, imposing this condition turns out to 
be \emph{sufficient} for existence of solutions to both (\ref{kseboundary}) and (\ref{KSE}). 
We will show that this is the case by providing the explicit solutions. 

Firstly, the condition (\ref{cool}) implies that the quartic 
polynomials $\mathcal{P}(p)$ and  $\mathcal{Q}(q)$
factorise as  $\mathcal{P}(p) =  w_-(p)w_+(p)$ and $\mathcal{Q}(q) =  w_-(q)w_+(q)$, 
where\footnote{Note that the coefficients in these quadratic functions may be complex, and we have taken the branch  of the square root 
$\sqrt{\alpha} = \tfrac{1}{2} (\tfrac{M^2}{Q^2} + E)$.} 
\bea
w_+ (x) & =& x^2 +\frac{M}{Q}x+Q + \sqrt{\alpha} ~,\nn\\
w_- (x) & = & x^2 -\frac{M}{Q}x-Q + \sqrt{\alpha}~.
\eea
Writing the  two-component  spinor as
\bea
\chi &= &\left(\begin{array}{c} \chi^+ \\ \chi^- 
\end{array} \right)~,
\eea
the integrability condition \eqref{integrbound12} implies that 
\bea
\chi^+ &=&  \sqrt{\frac{w_+(p)}{w_-(p)}} \, \chi^-~.
\label{goofy}
\eea
Using  this, it is  straightforward to find the general solution to equation (\ref{kseboundary}), which  in the frame (\ref{3dframe}) reads
\bea
\chi  &=& \left( \begin{array}{c} \sqrt{w_+(p)}\\[1.7mm] \sqrt{w_-(p)} \end{array} \right) \cdot   \exp{\left(\ii \tfrac{M}{2Q}(\tau+\sqrt{\alpha}\sigma)\right)} ~,
\label{singspinors}
 \eea
up to a complex constant. We shall discuss global properties of these spinors momentarily. Employing
the integrability condition (\ref{interel}) we have  determined the full solution to the four-dimensional Killing spinor equation 
(\ref{KSE}), which in the frame (\ref{4dframe}) reads
\bea
\epsilon_+ \ = \ 
\left(
\begin{array}{c}
-\ii \sqrt{\frac{w_+(q) }{q+p}}  \, \chi^+ \\
 -\ii \sqrt{\frac{w_-(q)}{q+p}}  \, \chi^- \end{array}\right)~, \qquad  
 \epsilon_-  \ =  \ 
\left(
\begin{array}{c}
 \sqrt{\frac{w_-(q)}{q-p}} \, \chi^+\\
 \sqrt{\frac{w_+(q)}{q-p}} \, \chi^- \end{array}\right)~.
\eea
Note that the expansions of these for large $q$ agree  with our initial asymptotic form of the spinors in (\ref{expaspinors}).

A convenient parameterisation of the solutions  to (\ref{cool}) can be obtained 
in terms of the four roots of the quartic polynomial $\mathcal{P}(p)$.
Specifically,  the condition (\ref{cool}) yields the following solutions:
\bea
Q &=& \begin{cases} \pm \frac{(p_3+p_1)(p_4+p_1)}{2}\\ 
\pm \frac{(p_3+p_4)(p_3+p_1)}{2} \\ \pm \frac{(p_3+p_4)(p_4+p_1)}{2}  \end{cases}.
\label{qcases}
\eea
Notice that although the roots $p_3,p_4$ are real by definition, the root $p_1$ may be complex, as discussed in section \ref{sugrasol}.
Hence $Q$ is generically \emph{complex}, implying the fields $A^{(3)}_\alpha$ and $V^{(3)}_\alpha$ can take complex values.

Let us now return to the spinors. In order  to check that these are globally defined on the three-sphere we have to change angular variables from $\tau,\sigma$ to 
$\varphi_1,\varphi_2$ as in section \ref{boundreg}. After doing so,  the argument of the exponential in general does not lead to a correct transformation of the spinors under shifts $\varphi_i \to \varphi_i + 2\pi$. 
However, this form of the spinors was obtained using the singular gauge field in (\ref{aandv}) 
(while the frame (\ref{3dframe}) is globally defined). If we instead use the globally defined gauge field 
\bea
A^{(3)}_\mathrm{global} & = & Q p \diff\sigma + \frac{Q}{p_3+p_4} (\diff \tau - p_3 p_4 \diff \sigma)~,
\eea
inherited from \eqref{globala}, the exponential factor in the Killing spinor (\ref{singspinors}) changes accordingly to 
\bea
 \exp{\left(\ii \left[\tfrac{M}{2Q}(p_3^2 -\sqrt{\alpha}) + Q p_3 \right]\phi_1 +\ii \left[\tfrac{M}{2Q}(p_4^2 -\sqrt{\alpha}) + Q p_4 \right]\phi_2 \right)}~.
\eea
Remarkably, using the explicit form of the solutions for $Q$ in (\ref{qcases}), we find that the expressions simplify and lead to the following
final form of the Killing spinors
\bea
\chi  (Q) &=& \left( \begin{array}{c} \sqrt{w_+(p)}\\[1.7mm] \sqrt{w_-(p)} \end{array} \right) \cdot  
\begin{cases}  \exp{\tfrac{\ii}{2}(\varphi_1 + \varphi_2)} ~~~~~~~~~~\mathrm{Type~I}\\[1.7mm] \exp{\tfrac{\ii}{2}(-\varphi_1 + \varphi_2)} \\[1.7mm]  \exp{ \tfrac{\ii}{2}(\varphi_1 - \varphi_2)}\end{cases} 
\begin{array}{l}
  \\[2.5mm]
 \!\!\!\!\!\!\!\! \!\!\!\!\!\!\!\! \!\!\!\!\!\!\!\!\!\!\!\mathrm{Type~II}\\
  \end{array},
\label{globalspinors}
 \eea
where we made a conventional choice of signs picking the  lower (minus) signs in \eqref{qcases}.
However,  since all solutions of $Q$ come in pairs with opposite signs, and the sign of $Q$ affects 
$A^{(3)}_\alpha$ and $V^{(3)}_\alpha$, but not  the function $H$ and the metric, for any solution  $\chi(Q)$ in \eqref{globalspinors}, 
we obtain also a spinor $\tilde \chi (Q)= \chi (-Q)$, solving the second equation in 
(\ref{mainrigid}). More details on the properties of the spinors may be found in appendix \ref{furtherapp}.

Finally, to see that \eqref{globalspinors} are globally defined spinors on the three-sphere recall from section 
\ref{boundreg} that near to the end-points  $p=p_3, p=p_4$, where  either $w_-(p)$ or $w_+(p)$ vanish, the three-sphere
looks like $\R^2\times S^1$, and indeed the phases in (\ref{globalspinors}) correspond to the correct anti-periodic spinors on $\R^2$.

\section{Parameter space}
\label{parameters}

Below we will discuss the change of coordinates and the choice of parameterisation that lead to the metric and free energies presented in section \ref{introsumma}.
%at the beginning of the  paper. 
Recall that a solution is specified by three parameters, for example the two real roots $p_3,p_4$, and a third\footnote{When this is complex, the third independent parameter is 
the imaginary part Im[$p_1$], while the real part is given by Re$[p_1]=-\tfrac{1}{2}(p_3+p_4)$.} root $p_1$. 
The scaling symmetry $p\to \lambda p$, $q\to \lambda q$ can be used to fix one of these parameters to any value, provided it is different from zero. Noting that $p_3+p_4>0$, 
we can define 
\bea
s \ =\ \tfrac{1}{2}(p_3+p_4) ~,\qquad a \ = \ \tfrac{1}{2}(p_4-p_3)   ~, 
\eea
and take $s,a,M$ as  real independent parameters.  The roots are given by $p_3=s-a$, $p_4=s+a$ and $p_1=-s - \sqrt{a^2 - \tfrac{M}{s}}$, 
$p_2=-s + \sqrt{a^2 - \tfrac{M}{s}}$. We can now set $s$ to a particular non-zero value, and without loss of generality we will set $s=\tfrac{1}{2}$. We then make the following change
of coordinate
\bea
p & = & \tfrac{1}{2} - a \cos\theta~,
\eea
where $\theta \in [0,\pi]$. Although in the original coordinate $p$ the parameter $a$ has to be strictly positive, after changing coordinates, the metric in the variables 
$\theta,\varphi_1,\varphi_2$ has a smooth limit  $a\to 0$. Indeed, precisely in this special case the metric reduces the Taub-NUT-AdS metric \cite{Martelli:2011fw}, whose 
boundary is the biaxially squashed metric (\ref{fametric}), with parameters identified as  $M = \tfrac{v^2-1}{2}$. Moreover, since
positive and negative values of  $a$ are simply related  by the change of coordinates $\theta \to \pi -\theta$, we can also take $a<0$.

When $p_2 \in \R$, from $p_2<p_3$ it follows immediately that 
\bea
2 M + 1 \ > \ 2 |a|\ > \ 0~.
\label{avwedge}
\eea
Moreover, when $p_2\in \C$, we have $a^2 - 2 M <0$, which again implies the inequality (\ref{avwedge}) holds. 
%{\color{red} The contraint $p_2 < p_3$ is not true in general since $p_2,p_3$ can be complex. However when $p_2, p_3$ are complex
%\bea
%2M + 1 &>& a^2 +1 = (a-1)^2 + 2a \Rightarrow 2M+1 > 2a~,~~~a>0  \nn \\
%2M + 1 &>& a^2 +1 = (a+1)^2 - 2a \Rightarrow 2M+1 > -2a~,~~~a<0 
%\eea
%and so $2 M + 1 \ > \ 2 |a|$ still holds.
%}
Therefore,  introducing the parameter $v^2=2M+1$ without loss of generality,  our solutions are parameterised by $a,v$, subject to the constraint 
$v^2 > 2 |a|$. The final form of the boundary metric and background fields are given in \eqref{finalmetric} and \eqref{finalbacks}. The bulk metric, gauge field, and 
Killing spinors in these coordinates and parameters, are not particularly simple and therefore we will not present them here. In terms of the parameters  $a$ and $v$,  
the solutions (\ref{qcases}) read
\bea
Q = \begin{cases}  \mp \frac{v^2-1}{2} \\[1.8mm] 
\mp \frac{1}{2} (\sqrt{a^2 +1 - v^2} +a)  \\[1.8mm]
\mp \frac{1}{2} (  \sqrt{a^2 +1 - v^2} -a)
  \end{cases},
\label{Qav}
\eea
from which it is manifest that $Q$ can take both real or complex values.

\begin{figure}[ht!]
\centering
\includegraphics[width=0.6\textwidth]{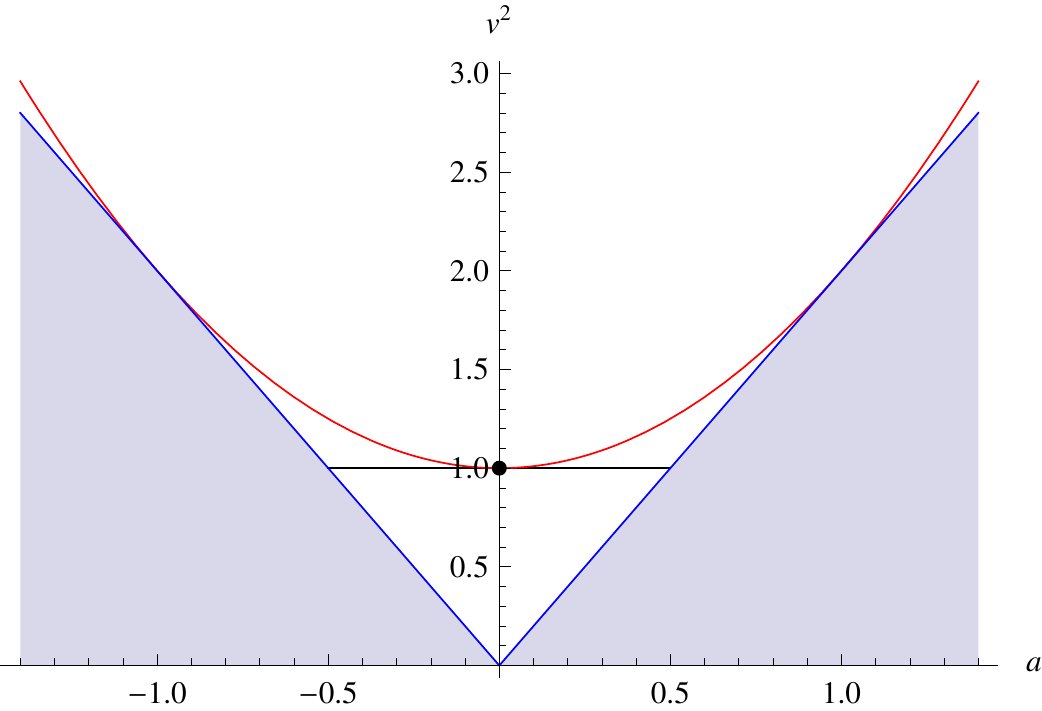}
\caption{Solutions exist for all parameters inside the wedge. Solutions for parameters below the parabola are always \emph{real}. 
Solutions for parameters above the parabola are \emph{complex} in the Type II cases. The black dot represents Euclidean AdS$_4$, with round three-sphere boundary.}\label{vapig}
\end{figure}

We have plotted the parameter space of solutions in the $(a,v^2)$ plane in Figure \ref{vapig}. The solutions exist inside the wedge defined 
by   $v^2 -2 |a| >0$. Although the metric is always real, in the Type II case the gauge field is complex for  values of the parameters above the 
 parabola\footnote{The parabola defines a locus where there is a double root $p_1=p_2$ of $\mathcal{P}(p)$ (which becomes a triple root $p_1=p_2=p_3=-\tfrac{1}{2}$ at $|a|=1$, $v^2=2$).  
 If $|a| > 1$  the double root $p_1>p_3$, which is not allowed (our regularity analysis in section \ref{sugrasol} does not apply). If $|a|  < 1$ the double root 
 $p_1<p_3$, which is allowed. Therefore the central arc of the parabola corresponds to a real solution for any choice of $Q$.} 
$v^2 =  a^2 +1$.

Two special loci correspond to the one-parameter solutions that appeared before in the literature. 
The $v^2$ axis, at $a=0$,  corresponds to the Taub-NUT-AdS solutions in \cite{Martelli:2011fw,Martelli:2012sz}.  In this case there are two inequivalent choices of $Q$ (up to signs), which
 correspond to the 1/4 BPS and 1/2 BPS instantons discussed in \cite{Martelli:2012sz}.  Notice the latter is real or pure imaginary depending on whether $v^2$ is smaller or larger than $1$.
  The segment at $v^2=1$, parameterised by $a\in [-\tfrac{1}{2},\tfrac{1}{2}]$, corresponds to the solution 
 of  \cite{Martelli:2011fu}. To see this, one has to identify the parameters as $a=\tfrac{1}{2} \tfrac{b^2-1}{b^2+1}$, 
 while the change coordinates between $\theta$ and $\vartheta$  may be obtained equating the respective functions $H(\theta)$ and $H(\vartheta)$.  
  In this case, two values of $Q$ in (\ref{Qav}) vanish, leaving only the (real) instanton discussed in 
 \cite{Martelli:2011fu}.

 \section{Holographic free energy}
\label{holofree}

In this section we compute the holographic free energy associated to our supergravity solutions 
using standard holographic renormalization methods \cite{Emparan:1999pm, Skenderis:2002wp}. 
The total on-shell action is
\bea
I &=&  I^{\text{grav}}_{\text{bulk}} + I^F + I^{\text{grav}}_{\text{ct}} + I^{\text{grav}}_{\text{bdry}}~.
\label{alli}
\eea
Here the first two terms are the bulk supergravity action (\ref{4dSUGRA}) 
\bea
I^{\text{grav}}_{\text{bulk}} + I^F &\equiv & -\frac{1}{16\pi G_4}\int \diff^4x\sqrt{g_{\mu\nu}}\left(R + 6 - F^2 \right)~,
\eea
evaluated on a particular solution. This is divergent, but we may regularize it using holographic renormalization. Introducing 
a cut-off at some large value of $q=\rcut$, with corresponding hypersurface $\mathcal{S}_\rcut
=\{q=\rcut\}$, we  add the following boundary terms
\bea
I^{\text{grav}}_{\text{ct}} + I^{\text{grav}}_{\text{bdry}} &=&  \frac{1}{8\pi G_4} \int_{\mathcal{S}_\rcut} \diff^3x \sqrt{\gamma_{\alpha\beta}} \left( 2 + \tfrac{1}{2} R(\gamma_{\alpha\beta}) - K\right) ~.
\eea
Here $R(\gamma)$ is the Ricci scalar of the induced metric $\gamma_{\mu\nu}$ on $\mathcal{S}_\rcut$, and $K$ is the trace of
 the second fundamental form of $\mathcal{S}_\rcut$, the latter being the Gibbons-Hawking boundary term. We compute
\bea
I^{\text{grav}}_{\text{bulk}} &=& \frac{\varpi}{8\pi G_4} \left[ p_3p_4(p_4^2-p_3^2)  - (p_4^3-p_3^3)\varrho + (p_4-p_3)\varrho^3 \right] ~, \\
I^{\text{grav}}_{\text{ct}} + I^{\text{grav}}_{\text{bdry}} &=&  \frac{\varpi}{8\pi G_4} \left[M(p_4-p_3)  - (p_4-p_3) \varrho^3 + (p_4^3-p_3^3) \varrho +O(\varrho^{-1}) \right] ~. \nn
\eea
As expected, the divergent terms cancel as $\rcut \to \infty$. In the above expressions we have introduced 
\bea
\varpi \ \equiv \ \int \diff\sigma \diff\tau \ =   \ - 16\pi^2 \frac{p_4^2-p_3^2}{\mathcal{P}'(p_3)\mathcal{P}'(p_4)}~,
\eea
where the integral is computed after writing $\tau, \sigma$ in terms of the coordinates  $\varphi_1,\varphi_2$.
The contribution to the action of the gauge field is finite in all cases and does not need regularization. In particular, we have
\bea
I_F &=& Q^2 \frac{\varpi}{8\pi G_4}\frac{p_4-p_3}{p_4+p_3}~.
\eea

Combining all the above contributions to the action we obtain the following expression:
\bea
I &=& \frac{\varpi}{8 \pi G_4}\left[p_3p_4(p_4^2-p_3^2)+ M(p_4-p_3) + Q^2 \frac{p_4-p_3}{p_4+p_3} \right]~.
\eea

Substituting for $M = -\frac{1}{2}(p_3+p_4)(p_3p_4-p_1p_2)$ and $Q$ given in (\ref{qcases}), we find the following values of the free energies
\bea
I &=& \frac{\pi}{2 G_4} \begin{cases}   1\\ \frac{(p_3+p_4)^2}{(p_4-p_1)(p_3-p_2)} \\  \frac{(p_3+p_4)^2}{(p_3-p_1)(p_4-p_2)} \end{cases}~,
\eea
in the three cases, respectively. 
Finally, writing this  in terms of the parameters $a$ and $v$, we obtain
\bea
I = \frac{\pi}{2G_4}
\begin{cases} 1  \qquad \qquad \mathrm{Type~I}\\[2mm]
 \frac{1}{ 1  - 4Q^2}  \qquad \mathrm{Type~II}
\end{cases} .
\label{qisfree}
\eea

Remarkably, the free energy can  always be expressed entirely in terms  of $Q^2$.
After the  change of  variables $Q   =  \frac{1}{2}\frac{\beta^2-1}{\beta^2+1}$,
the free energy in the Type II case takes the familiar form 
\bea
I = \frac{\pi}{8G_4} \left( \beta+ \frac{1}{\beta} \right)^2~,
\eea
where, definig $a  = \frac{1}{2}\frac{b^2-1}{b^2+1}$,  the parameter $\beta $ is given  by 
\bea
\beta^2 & = &   \frac{(1+b^2)(2-v^2) + \sqrt{(b^2-1)^2- 4 (v^2-1) (b^2+1)^2}}{1 -b^2  + v^2 + b^2 v^2}~.
\label{thisisbeta}
\eea
When $v=1$ we have  $b=\beta$, while  when $b=1$, we have $\tfrac{2}{v}= \beta + \tfrac{1}{\beta}$, and one can check that 
the free energies reduce to  those computed previously in the special cases.

\section{Discussion}
\label{discussect}

In this paper we constructed  a family of rigid supersymmetric geometries   depending on two parameters, comprising a deformed three-sphere and various background fields. 
These interpolate between all the previously known rigid supersymmetric geometries with topology of 
the three-sphere \cite{Hama:2011ea,Imamura:2011wg}. ${\cal N}=2$ supersymmetric gauge theories may be placed on these backgrounds, with precise
Lagrangians and supersymmetry transformation rules \cite{Klare:2012gn,Closset:2012ru}, and we have presented  supergravity solutions conjecturally dual to these. 
Although these were obtained in $d=4$, ${\cal N}=2$ gauged supergravity, using the results of 
\cite{Gauntlett:2007ma} and \cite{Martelli:2012sz}, all our solutions may be uplifted to global supersymmetric solutions of (Euclidean) M-theory.
We have  computed the holographic free energy in the various cases,  finding that it is either constant, or it depends on the parameters in a simple way, 
thus making a prediction for the large $N$ limit of the localized partition function of a large class of supersymmetric guage 
theories. This strongly suggests that  the \emph{full} localized partition function on these backgrounds 
may be written  in terms  of the double sine function $s_\beta (x)$, where in the Type I solutions $\beta=1$, while in the Type II solutions $\beta$ is given in \eqref{thisisbeta}.
 More generally, it suggests that on any supersymmetric geometry with $S^3$ topology, the partition function 
 can be expressed in terms of $s_\beta (x)$, for an appropriate $\beta$.  It would be interesting to understand better the geometric interpretation\footnote{After the first version of 
 this paper was submitted on the arXiv, in \cite{Alday:2013lba} our conjectured form of the localized partition function has been proved, and the significance of $\beta$ has been elucidated.} of this $\beta$.

\subsection*{Acknowledgments}
We would like to thank James Sparks for collaboration at the initial stages of this work.
D. M. is supported by an ERC Starting Grant -- Grant Agreement N. 304806 --  Gauge-Gravity,
and also acknowledges partial support from an EPSRC Advanced Fellowship EP/D07150X/3 and from the STFC grant ST/J002798/1. 
A.P. is supported by an A.G. Leventis Foundation grant,  an STFC studentship and via the Act ``Scholarship Programme of S.S.F. by 
the procedure of individual assessment, of 2011-12'' by resources of 
the Operational Programme for Education and Lifelong Learning, of the European Social Fund (ESF) 
and of the NSRF, 2007-2013.

\appendix

\section{Integrability conditions}
\label{integrab}

\subsection{Integrability condition of the bulk Killing spinor equation}

The integrability condition of \eqref{KSE} reads:
\bea\label{integrbulk}
\frac{1}{4}R_{\mu\nu}{}^{ab}\Gamma_{ab}\epsilon + \frac{1}{2}\Gamma_{\mu\nu}\epsilon &=& 
\ii F_{\mu\nu}\epsilon - \frac{\ii}{2}\nabla_{[\mu }F_{|\, ab}\Gamma^{ab}_{\ \ |}\Gamma_{\nu]}\epsilon - \frac{\ii}{4}\Gamma_{[\mu }F_{|\, ab}\Gamma^{ab}_{\ \ |}\Gamma_{\nu]}\epsilon\nonumber\\
&& + \frac{1}{16}\left[F_{ab}\Gamma^{ab}\Gamma_\mu,F_{cd}\Gamma^{cd}\Gamma_\nu\right]\epsilon -\frac{\ii}{4}F_{ab}\Gamma^{ab}\Gamma_{\mu\nu}\epsilon~.
\eea
We shall use the above equation to obtain an algebraic relation between $\epsilon_+$ and $\epsilon_-$. We need to extract only one non-trivial component.
In the orthonormal frame  \eqref{4dframe} we pick the $23$ component. Using
\bea
R_{2314} \ = \ \frac{M}{(p+q)^3}~,~~~~~R_{2323} \ = \ -1-\frac{M}{(p+q)^3}
\eea
and
\bea
\nabla_3 F_{14} \ = \ \nabla_3 F_{32} &=& \frac{Q}{(p+q)^3} \sqrt{\frac{\mathcal{P}(p)}{p^2-q^2}} \nn \\
\nabla_2 F_{14} \ = \ \nabla_2 F_{32} &=& - \frac{Q}{(p+q)^3} \sqrt{\frac{\mathcal{Q}(q)}{q^2-p^2}}
\eea
we derive 
\bea\label{interel}
\epsilon_+ =  \Omega \; \epsilon_- ~,~~~~~
\Omega \ = \ Q \begin{pmatrix} \frac{\ii \sqrt{\frac{\mathcal{Q}(q)}{q^2-p^2}}}{M-(p+q)Q} & \frac{\sqrt{\frac{\mathcal{P}(p)}{p^2-q^2}}}{M-(p+q)Q} \\ \frac{-\sqrt{\frac{\mathcal{P}(p)}{p^2-q^2}}} {M+(p+q)Q} & \frac{-\ii \sqrt{\frac{\mathcal{Q}(q)}{q^2-p^2}}}{M+(p+q)Q} \end{pmatrix}~.
\label{omegamatrix}
\eea

\subsection{Integrability condition of the boundary Killing spinor equation}

The integrability condition of \eqref{kseboundary} reads\footnote{The first version of this paper contained a sign error 
in equation (\ref{integrbound}), that has been corrected in \cite{Alday:2013lba}.}:
\bea\label{integrbound}
\biggl[ \frac{1}{4} R^{(3)}_{\alpha\beta\delta\epsilon} \gamma^{\delta\epsilon} - \ii F^{(3)}_{\alpha\beta} + \ii \partial_{[\alpha} p \gamma_{\beta]} - \frac{1}{2} p^2 \gamma_{\alpha\beta}
- 2 \ii \nabla^{(3)}_{[\alpha|} V^{(3)}_\delta \gamma_{|\beta]} \gamma^\delta \nn \\
+ 2 p \gamma_{[\alpha} V^{(3)}_{\beta]} + 2V^{(3)}{}^\delta V^{(3)}_\delta \gamma_{\alpha\beta} -4 V^{(3)}_\delta \gamma_{[\alpha} V^{(3)}_{\beta]} \gamma^\delta \biggr] \, \chi &=& 0~.
\eea
In the orthonormal frame  \eqref{3dframe} we pick the $12$ component. Using
\bea
R^{(3)}_{1213} &=&  - \sqrt{-\mathcal{P}(p)} \nn \\
R^{(3)}_{1212} &=& 3p^2 + E \nn \\ 
\nabla_1^{(3)} V^{(3)}_2 \ = \   - \nabla_2^{(3)} V^{(3)}_1 &=& \frac{M}{2Q} p
\eea
we find
\bea\label{integrbound12}
\Omega^{(3)}  \chi &=& 0
\eea
where
\bea
\Omega^{(3)} \ = \ 
\begin{pmatrix}
p^2 - \frac{M}{Q}p - Q + \frac{1}{2}\left(\frac{M^2}{Q^2}+E\right) & -\ii\sqrt{-\mathcal{P}(p)} \\ 
\ii\sqrt{-\mathcal{P}(p)} & -p^2 - \frac{M}{Q}p  -Q -\frac{1}{2}\left(\frac{M^2}{Q^2}+E\right)
\end{pmatrix}~.
\eea
In order for \eqref{integrbound12}  to have a solution, the determinant
\bea
\det \Omega^{(3)} &=& \frac{1}{4}\left(\frac{M^2}{Q^2}+E\right)^2 - \alpha 
\eea
must be zero. Using the relations
\bea
E &=& p_1p_2 +p_1p_3+p_2p_3+p_1p_4+p_2p_4+p_3p_4 \nn \\
M &=& \tfrac{1}{2}(p_1p_2p_3+p_1p_2p_4+p_1p_3p_4+p_2p_3p_4) \nn \\
\alpha &=& p_1p_2p_3p_4 +Q^2 \nn \\
0 &=& p_1+p_2+p_3+p_4
\eea
we derive
\bea
\det \Omega^{(3)} &=& \frac{\alpha_1\alpha_2\alpha_3}{(8Q^2)^2}~,
\eea
where
\bea
\alpha_1 &=& (p_3+p_4)^2(p_3+p_1)^2-4Q^2 \nn \\
\alpha_2 &=& (p_3+p_4)^2(p_4+p_1)^2-4Q^2 \nn \\
\alpha_3 &=& (p_3+p_1)^2(p_4+p_1)^2-4Q^2~.
\eea
Hence we obtain the following possibilities:
\bea
\label{onemore}
Q = \begin{cases} \pm \frac{(p_3+p_1)(p_4+p_1)}{2}\\ \pm \frac{(p_3+p_4)(p_3+p_1)}{2} \\ \pm \frac{(p_3+p_4)(p_4+p_1)}{2} \end{cases}~.
\eea
In addition, \eqref{integrbound12} relates $\chi^-$ and $\chi^+$ as
\bea
\chi^+ &=&  \sqrt{\frac{p^2 +\frac{M}{Q}p+Q + \sqrt{\alpha}}{p^2 -\frac{M}{Q}p-Q + \sqrt{\alpha}}} \, \chi^-~,
\eea
which is equation (\ref{goofy}) in the main text.

\section{More on the Killing spinors}

\label{furtherapp}

In this appendix we discuss  properties of the Killing spinors and some of their bilinears. 

\subsection{$\chi$, $\chi^c$ and $\tilde \chi$}

We may write down the following three, in general distinct, Killing spinor equations:
\bea
\label{rigid1}
\nabla_\alpha^{(3)} \chi- \ii (A^{(3)}_\alpha  + V^{(3)}_\alpha)\chi + \tfrac{1}{2}H \gamma_\alpha \chi + \epsilon_{\alpha\beta\rho}V^{(3)}{}^\beta\gamma^\rho\chi  &=&  0 ~,\\[2mm]
\label{rigid2}\nabla_\alpha^{(3)} \chi^c + \ii (\bar A^{(3)}_\alpha  + \bar V^{(3)}_\alpha)\chi^c - \tfrac{1}{2}\bar H \gamma_\alpha \chi^c - \epsilon_{\alpha\beta\rho}\bar V^{(3)}{}^\beta\gamma^\rho\chi^c  &=&  0 ~,\\[2mm]
\label{rigid3}\nabla_\alpha^{(3)} \tilde \chi + \ii (A^{(3)}_\alpha  + V^{(3)}_\alpha)\tilde \chi + \tfrac{1}{2}H \gamma_\alpha \tilde \chi -\epsilon_{\alpha\beta\rho}V^{(3)}{}^\beta\gamma^\rho\tilde \chi & = & 0 ~.
\eea
Equation \eqref{rigid2} is the charge conjugate of  equation \eqref{rigid1}, where a bar in $\bar A^{(3)}$,   $\bar V^{(3)}$, $\bar H$ denotes complex conjugation and the charge conjugate spinor $\chi^c$ is  defined as
\bea
\chi^c \ = \ C \chi^*~,~~~~~~~ C \ = \ \begin{pmatrix} 0 &1 \\ -1 & 0 \end{pmatrix}~.
\eea
Notice that \eqref{rigid2} is also obtained from \eqref{rigid1} by replacing 
\bea
 A^{(3)}~\to ~ - \bar A^{(3)}~, \qquad    V^{(3)} ~\to ~   -\bar V^{(3)} ~, \qquad H ~ \to ~ - \bar H ~,
 \label{getconjugate}
\eea
and for any solution $\chi$ of \eqref{rigid1}, $\chi^c$ is a solution of \eqref{rigid2}. On the other hand, \eqref{rigid3} is obtained from \eqref{rigid1} by replacing
\bea
 A^{(3)} ~\to ~ - A^{(3)}~, \qquad    V^{(3)} ~\to ~   - V^{(3)} ~, \qquad H ~ \to ~  H ~, 
 \label{getilde}
\eea
and in general is an independent equation. In particular, the existence of a  solution $\chi$ to \eqref{rigid1} does \emph{not} imply that there exists a
solution $\tilde \chi$ to \eqref{rigid3}. 
There are two special cases:
\begin{enumerate}
\item if   $A^{(3)}$, $V^{(3)}$ are \emph{real} and   $H$ is pure imaginary
then \eqref{rigid2} and \eqref{rigid3} coincide. In this case, for any solution $\chi$ to \eqref{rigid1} there is also a solution  $\tilde\chi=\chi^c$ to \eqref{rigid3}.

\item if   $A^{(3)}$, $V^{(3)}$   and   $H$ are  pure \emph{imaginary} then \eqref{rigid1} and \eqref{rigid2} coincide. In this case,
for any solution $\chi$ to \eqref{rigid1} there is also a second solution 
$\chi^c$ to \eqref{rigid1}. As in Euclidean signature $\chi$  and $\chi^c$ are independent, this implies that these configurations are 1/2 BPS. 

\end{enumerate}

Let us now discuss how our solutions fit into these relationships. We chose conventionally to refer to the solutions with a specific choice of signs of $Q$ (lower signs in \eqref{onemore})
as spinors $\chi$ solving \eqref{rigid1}. Namely, we take 
\bea
\chi (Q)  \ = \ \begin{pmatrix} \sqrt{w_+(p)}  \\ \sqrt{w_-(p)} \end{pmatrix} \ex^{\ii\Phi (Q)}  ~,
\label{basicspinor}
\eea
where 
\bea
\label{blabla}
Q=
\begin{cases} 
 \frac{v^2-1}{2} \\[1.8mm]
 \frac{1}{2} (\sqrt{a^2 +1 - v^2} +a)  \\[1.8mm]
 \frac{1}{2} (  \sqrt{a^2 +1 - v^2} -a)
 \end{cases},
\qquad\quad  
\Phi (Q)  \ = \ \begin{cases} \frac{1}{2}(\varphi_1+\varphi_2) \\[1.8mm] \frac{1}{2}(-\varphi_1 + \varphi_2) \\[1.8mm] \frac{1}{2}(\varphi_1 -\varphi_2)  \end{cases}.
\eea
Then using the fact that under $Q\rightarrow -Q$ our background fields  transform as in (\ref{getilde}) and 
$w_+ (p) \leftrightarrow w_-(p)$, $\Phi \rightarrow -\Phi$, 
it follows that
\bea
\tilde \chi (Q) \ \equiv  \ \chi(-Q) \ = \ \begin{pmatrix} \sqrt{w_-(p)}  \\ \sqrt{w_+(p)} \end{pmatrix} \ex^{-\ii\Phi (Q) }~,
\label{tildespin}
\eea
is a solution to \eqref{rigid3}, for all choices of $Q$ in \eqref{blabla}. 

Let us now look at the charge conjugate of \eqref{basicspinor}. In general this reads
\bea
\chi^c (Q) \  = \ \begin{pmatrix} ~\left(\sqrt{w_-(p)}\right)^*  \\[1.8mm] -\left(\sqrt{w_+(p)}\right)^* \end{pmatrix} \ex^{-\ii\Phi(Q)}~.
\label{goofy}
\eea
Equation \eqref{rigid2} can be obtained from \eqref{rigid1} transforming the fields as in \eqref{getconjugate}, which in our solutions
corresponds to 
\bea
Q \ \to \  - Q^*~. 
\label{recharge}
\eea
Therefore we should find that under \eqref{recharge} the spinor $\chi (Q) \to \chi^c(Q)$. Let us 
first assume that $Q\in \R$, then  $\chi(Q) \to \tilde \chi (Q)$ as in \eqref{tildespin} and we have
\bea
\chi^c (Q) \  = \ \begin{pmatrix} ~\left(\sqrt{w_-(p)}\right)^*  \\[1.8mm] -\left(\sqrt{w_+(p)}\right)^* \end{pmatrix} \ex^{-\ii\Phi(Q)}\ = \
\pm \begin{pmatrix} \sqrt{w_-(p)}  \\ \sqrt{w_+(p)} \end{pmatrix} \ex^{-\ii\Phi (Q) } = \pm \tilde \chi (Q)  ~.
\eea
Here we have used the fact that for $Q \in \mathbb{R}$,  $w_-(p)$ and $w_+(p)$ are real and  
since  $w_-(p)  w_+(p) = \mathcal{P}(p) \leq 0 $ it follows that either $w_-(p)\leq 0$, $w_+(p)\geq 0$ or  $w_-(p)\geq 0$, $w_+(p)\leq 0$, resulting in the two signs above.

More generally, when $Q\in \C$, \eqref{recharge} implies that
\bea
\chi (Q) \ \to \ \chi (-Q^*) = \begin{pmatrix} \sqrt{w_-(p)^*}  \\ \sqrt{w_+(p)^*} \end{pmatrix} \ex^{-\ii\Phi (Q)} ~,
\label{gocompare}
\eea
where we used the fact that the two (lower) Type II cases of $Q$ in \eqref{blabla} are exchanged under \eqref{recharge}. In order to compare \eqref{gocompare} to
\eqref{goofy} we must use the fact that 
\bea
w_+(p) = |w_+(p)|\mathrm{e}^{\ii\varphi_+ (p)} ~, ~ w_-(p) = |w_-(p)|\mathrm{e}^{\ii\varphi_- (p)} ~, ~~\mathrm{with}~~  \varphi_+ (p) + \varphi_- (p) = \pi~.
\eea
One can then check that in order for \eqref{gocompare} to agree (up to a possible sign depending on the signs of Re$[w_\pm(p)]$) with \eqref{goofy} one should define the square root on 
the complex plane with  a branch cut along the positive imaginary axis\footnote{We thank Nikolay Gromov for suggesting this.}. This definition would fail to give the correct relation
for purely imaginary values of $w_\pm (p)$, but this of course cannot happen.

We have therefore shown that the two  solutions of Type II are just charge conjugate to each other. 
A special case arises when $Q$ is purely imaginary, which we discuss below.

\subsection{Coordinate $\theta$ and the $a=0$ case}

To write the spinors in the coordinate $\theta$ one should make 
the coordinate transformation $p = \frac{1}{2} - a \cos\theta$ and then 
substitute into \eqref{basicspinor} the following
\bea
\left\{\begin{array}{ccl} w_+(\theta) & = & - a^2 \sin^2\theta 
 \\ w_-(\theta) & = &  (a\cos\theta-1)^2  + 2Q -a^2 
\end{array}\right.
~~\mathrm{for}~ Q =  \frac{v^2-1}{2}~,
\eea
in the Type I solutions and
\bea
w_\pm(\theta) \ = \ 
\begin{cases} 
a (\cos\theta \pm 1)(a\cos\theta -1 \mp 2Q \pm a)   \\
a (\cos\theta \mp 1)(a\cos\theta -1 \mp 2Q \mp a) \\ 
\end{cases}
~~\mathrm{for}~
Q \ = \
\begin{cases}
\tfrac{1}{2}(\sqrt{a+1-v^2} + a) \\
\tfrac{1}{2}(\sqrt{a+1-v^2} - a)
\end{cases}
\eea
in the type II solutions.
Although the resulting expressions are not particularly simple, it is now possible to take the limit $a\to 0$. The Type I spinor then reduces to 
\bea
\chi_{a=0}   \ = \ \begin{pmatrix} 0  \\ v \end{pmatrix} \ex^{\tfrac{\ii}{2}(\varphi_1+\varphi_2)}  ~,  
\eea
that is the spinor of the 1/4 BPS biaxially squashed three-sphere \cite{Martelli:2012sz}. 
Upon rescaling  $w_\pm (\theta) \rightarrow  w_\pm (\theta)/a$ before taking the $a\to 0$, the Type II spinors instead reduce to 
\bea
\chi_{a=0} \ = \ \begin{pmatrix}  \cos\frac{\theta}{2} \\ \ii \sin\frac{\theta}{2} \end{pmatrix} \ex^{\tfrac{\ii}{2}(-\varphi_1+\varphi_2)}  ~, \qquad 
\chi^c_{a=0} \ = \ \begin{pmatrix}  \ii \sin\frac{\theta}{2} \\ \cos\frac{\theta}{2} \end{pmatrix} \ex^{\tfrac{\ii}{2}(\varphi_1-\varphi_2)} ~, 
\eea
up to  irrelevant constants. These are the two spinors of the 1/2 BPS biaxially squashed three-sphere \cite{Martelli:2012sz}. 
Notice that indeed they are charge conjugate to each other.  Moreover, when $v^2>1$, $Q$ is pure imaginary, and it follows from the discussion above that they are  both solutions 
to \eqref{rigid1}.

\subsection{Spinor bilinears}

We evaluate the bilinears $K_\alpha = \chi \gamma_\alpha \tilde \chi$ and $\rho_\alpha \ = \ \chi \gamma_\alpha \chi$
appearing in \cite{Closset:2012ru}\footnote{$\rho$ is denoted $P$ in  \cite{Closset:2012ru}.} for our solutions. 
The contraction of two spinors is defined as
\bea
\psi \zeta \ = \ C^{\alpha\beta} \psi_\beta \zeta_\alpha~.
\eea
We have
\bea
K & = & 2 \left[(p^2+\sqrt{\alpha})(p_3^2-p^2) + \mathcal{P}(p)\right]\diff\phi_1  + 2 \left[(p^2+\sqrt{\alpha})(p_4^2-p^2) + \mathcal{P}(p)\right]\diff\phi_2 ~,\nn \\
\rho & = &  \frac{-2\ex^{2\ii\Phi}}{\sqrt{-\mathcal{P}(p)}} 
\left[(\tfrac{M}{Q}p+Q) \diff p + \ii \, \mathcal{P}(p) [(p_3^2 - \sqrt{\alpha})\diff\phi_1+ (p_4^2 - \sqrt{\alpha})\diff\phi_2]\right]~, 
\eea
where recall that $\phi_1 = \frac{2}{\mathcal{P}'(p_3)} \varphi_1$ and $\phi_2 = \frac{2}{\mathcal{P}'(p_4)} \varphi_2$. 
The dual Killing vector field is 
\bea
K^\sharp \ = \ \frac{\sqrt{\alpha}+p_4^2}{p_3^2-p_4^2} \mathcal{P}'(p_3)  \partial_{\varphi_1} - \frac{\sqrt{\alpha}+p_3^2}{p_3^2-p_4^2}\mathcal{P}'(p_4)\partial_{\varphi_2} ~.
\eea
Notice that $K$, $K^\sharp$ are in general \emph{complex}. They become real if and only if
$\sqrt{\alpha}\in \R$, which can happen only when $Q$ is purely real or imaginary, as for the special cases previously 
studied in the literature. Both $K$ and $\rho$ are globally defined one-forms on the three-sphere. 
Furthermore, $\rho$ satisfies the condition  \cite{Klare:2012gn,Closset:2012ru} $\rho \wedge \diff \rho =  0$.

\end{document}